\documentclass[fleqn,usenatbib]{mnras}

\usepackage{newtxtext,newtxmath}

\usepackage[T1]{fontenc}

\DeclareRobustCommand{\VAN}[3]{#2}
\let\VANthebibliography\thebibliography
\def\thebibliography{\DeclareRobustCommand{\VAN}[3]{##3}\VANthebibliography}

\usepackage{graphicx}	
\usepackage{amsmath}	
\usepackage{nicefrac,xfrac}

\title[Quality flags for Gaia DR3 effective temperatures]{Quality flags for GSP-Phot Gaia DR3 astrophysical parameters with machine learning: Effective temperatures case study}

\author[A. S. Avdeeva et al.]{
Aleksandra S. Avdeeva,$^{1, 2}$\thanks{E-mail: avdeeva@inasan.ru}
Dana A. Kovaleva,$^{1}$
Oleg Yu. Malkov$^{1}$
and Gang Zhao$^{3}$
\\
$^{1}$Institute of Astronomy Russian Academy of Sciences, 48 Pyatnitskaya St., Moscow, 119017, Russia\\
$^{2}$HSE University, 20 Myasnitskaya St., Moscow, 101000, Russia \\
$^{3}$CAS Key Laboratory of Optical Astronomy, National Astronomical Observatories, Chinese Academy of Sciences Beijing 100101, China
}

\date{Accepted XXX. Received YYY; in original form ZZZ}

\pubyear{2015}

\begin{document}
\label{firstpage}
\pagerange{\pageref{firstpage}--\pageref{lastpage}}
\maketitle

\begin{abstract}

Gaia Data Release 3 (DR3) provides extensive information on the astrophysical properties of stars, such as effective temperature, surface gravity, metallicity, and luminosity, for over 470 million objects. However, as Gaia's stellar parameters in GSP-Phot module are derived through model-dependent methods and indirect measurements, it can lead to additional systematic errors in the derived parameters. In this study, we compare GSP-Phot effective temperature estimates with two high-resolution and high signal-to-noise  spectroscopic catalogues: APOGEE DR17 and GALAH DR3, aiming to assess the reliability of Gaia's temperatures. We introduce an approach to distinguish good-quality Gaia DR3 effective temperatures using machine-learning methods such as XGBoost, CatBoost and LightGBM. The models create quality flags, which can help one to distinguish good-quality GSP-Phot effective temperatures. We test our models on three independent datasets, including PASTEL, a compilation of spectroscopically derived stellar parameters from different high-resolution studies. The results of the test suggest that with these models it is possible to filter effective temperatures as accurate as 250~K with $\sim 90$ per cent precision even in complex regions, such as the Galactic plane. Consequently, the models developed herein offer a valuable quality assessment tool for GSP-Phot effective temperatures in Gaia DR3. Consequently, the developed models offer a valuable quality assessment tool for GSP-Phot effective temperatures in Gaia DR3. The dataset with flags for all GSP-Phot effective temperature estimates, is publicly available, as are the models themselves.

\end{abstract}




\section{Introduction}

Gaia Data Release 3 (DR3) \citep{Collaboration2016, Collaboration2022} represents a remarkable milestone in our understanding of the Milky Way and beyond, providing an unprecedented volume of precise astrometric, photometric, and stellar parameter data that will revolutionise various fields of astronomy. It includes information on astrophysical characteristics for hundreds of millions of stars obtained via several independent pipelines that have different sets of input data from Gaia observables \citep{Creevey2022}. Among the new products of Gaia DR3 one of the most awaited are the low-resolution spectra from the Blue (BP) and Red (RP) Photometer (the description and the internal calibration \citet{2021A&A...652A..86C, 2023A&A...674A...2D}, the external calibration \citet{2023A&A...674A...3M}), which allows for determination of astrophysical parameters for the hundreds of millions of stars.

One of the main modules, the General Stellar Parametrizer from photometry (GSP-Phot), estimates effective temperature $T_{\rm eff}$, logarithm of surface gravity $\textrm{log}g$, metallicity, absolute magnitude $\textrm{M}_G$, radius, distance, line-of-sight extinctions $\textrm{A}_\textrm{0}$, $\textrm{A}_G$, $\textrm{A}_{BP}$, and $\textrm{A}_{RP}$, and the reddening $E(BP-RP)$ by forward modelling the low-resolution BP/RP spectra, apparent G magnitude, and parallax using a Markov Chain Monte Carlo (MCMC) method. To this end, GSP-Phot employs stellar evolutionary models in forward model interpolation to obtain self-consistent temperatures, surface gravities, metallicities, radii, and absolute magnitudes \citep{Andrae2022}. The GSP-Phot module provides a range of astrophysical parameters resulting from different codes for atmosphere calculation, namely, MARCS \citep{2008A&A...486..951G}, PHOENIX \citep{2005ESASP.576..565B}, A and OB \citep{2003ApJS..146..417L,2007ApJS..169...83L} models. Gaia DR3 provides the best among the  stellar parameters obtained by the models with a GSP-Phot module for 471 million sources. Hereafter, we will consistently refer to these best-estimated parameters from GSP-Phot. 

These parameters facilitate in-depth studies of stellar properties, stellar evolution, and the composition of various stellar populations across the Galaxy. However, the self-consistent determination of these parameters may introduce additional systematic errors, particularly in the estimation of  $T_{\rm eff}$. In their study, \citet{2023ApJS..266...11B} utilized Gaia DR3 data as literature values for comparing their spectra calibration. To achieve this, they had to exclude all objects with  $T^{\rm Gaia}_{\rm eff}>7000~\rm K$ due to observed systematic differences. The work by \citet{2023arXiv230603132B} also addresses the discrepancy between Gaia DR3 GSP-Phot estimates of  $T_{\rm eff}$  and their own estimations for stars in the Hyades and Pleiades open clusters. Unfortunately, the GSP-Phot module, despite being the richest source of astrophysical parameters in Gaia DR3, does not provide quality flags or indicators upon which one can rely while using effective temperatures and other atmospheric parameters.

On the other hand, accurate determination of stellar temperatures is of paramount importance in various fields of astronomy and astrophysics. Knowledge of the precise and reliable temperatures of celestial objects enables us to unravel their intrinsic properties, understand their evolutionary stages, and gain insights into fundamental astrophysical processes. The effective temperatures from different spectroscopic surveys helped characterising the AGBb in the wide range of mass and metallicity \citep{2022A&A...668A.115D} and in providing detailed insights into disk/halo stars \citep{2021ApJ...916...88G}. Spectroscopic surveys also contribute to uncovering the characteristics of interstellar matter. The effective temperatures obtained from surveys such as LAMOST \citep{2015RAA....15.1095L} and RAVE \citep{McMillan2020} were utilized to calculate intrinsic colors in the works of \cite{2021EPJST.230.2193N} and \cite{2021OAst...30..168A}.  Based on these intrinsic colours, the interstellar visual extinction $\textrm{A}_{V}$ in several distinct lines of sight was calculated. Similarly, \citet{2021ApJS..254...38S} obtained interstellar
extinctions in the GALEX UV bands for over a million of stars using spectroscopic parameters from LAMOST and GALAH \citep{2021MNRAS.506..150B} surveys.  

At the moment there are several spectroscopic surveys with a decent resolution that could be considered a reliable source of fundamental stellar parameters, such as APOGEE \citep{Joensson2020, 2022ApJS..259...35A}, GALAH, Gaia-ESO \citep{Gilmore2012}, etc. APOGEE survey (DR17) $(\textrm{R} \sim 22500)$ covers a decent amount of areas in the Northern sky, with some parts in the Southern sky. APOGEE primarily aims to study evolved stars across the Galactic disc, the Galactic Centre, and the outer halo. A baseline magnitude limit of $H = 12.2$~mag was adopted for "normal" APOGEE fields. 
GALAH survey (DR3) $(\textrm{R} \sim 28000)$ explores the stars with the following magnitude selection: $12<V<14$, and the galactic latitude $|\textrm{b}|>10$ deg. 
Gaia-ESO survey $(\textrm{R} \sim 16000-25000)$ is a public spectroscopic survey that aims to obtain high-quality spectroscopy of 100000 Milky Way stars, systematically covering all major components of the Galaxy. Although these surveys have sufficient resolution and signal-to-noise ratio, they are limited in sky coverage and brightness of the stars. 

Efforts are being made to re-estimate the effective temperatures provided by Gaia DR3. In a recent study by \citet{2023ApJS..267....8A}, the authors employed a machine learning technique, specifically XGBoost, to estimate the effective temperatures, as well as metallicity and surface gravity. Machine learning model was trained on stellar parameters from APOGEE survey to predict the parameters from BP/RP spectra of Gaia DR3 and CatWISE magnitudes. The results are 175 million stars with re-estimated parameters, showing good agreement with APOGEE survey. \citet{2023MNRAS.524.1855Z} offer revised stellar atmospheric parameters for 220 million stars from Gaia DR3. Their approach employs a data-driven model of Gaia BP/RP spectra, trained using LAMOST data and augmented with 2MASS and WISE photometry. This method enhances precision and reduces parameter degeneracy, resulting in improved atmospheric parameter estimations.

The objective of the study is to comprehensively assess and compare effective temperature estimates derived in Gaia Data Release 3 (DR3) with those obtained from high-resolution spectroscopic surveys, specifically APOGEE and GALAH. The study aims to understand the level of agreement and discrepancies between these temperature estimates, considering various parameters and potential systematic effects. Most importantly, the study seeks to leverage machine learning techniques to distinguish good quality effective temperatures from Gaia DR3 data, making a tool that could provide the idea of whether the effective temperature for a particular object can be trusted and to what extent. 

This article is organised as follows. Section 2 compares Gaia temperatures with those from the APOGEE and GALAH surveys and indicates the regions in the parameter space where the differences are maximum/minimum. The data and methods for the machine learning approach are described in Section 3. Section 4 contains results of our study and we draw our conclusions in Section 5.

\section{Comparative analysis of effective temperatures: Gaia DR3 versus APOGEE and GALAH}
\label{sec:sec2}

In this section we compare the effective temperatures of Gaia DR3 provided by GSP-Phot module with the effective temperatures from APOGEE and GALAH using the stars Gaia DR3 and those catalogues have in common. Both APOGEE and GALAH benefit from a precalculated cross-match with the Gaia DR3 catalogue, which serves as a foundation for our analysis.

For clearer reference on the APOGEE/GALAH side, we use quality flags, recommended for these catalogues. For APOGEE, the selection criteria encompass $\textsc{aspcapflag} \neq \textsc{star\_bad}$, ensuring the use of the highest quality measurements with a favourable signal-to-noise ratio and the absence of pipeline-related issues. For GALAH, the criteria involve $\textsc{flag\_sp} = 0$, $\textsc{flag\_fe\_h} = 0$, and $\textsc{snr\_c3\_iraf} > 30$. These selection parameters are meticulously chosen to ensure the inclusion of data points that satisfy specific quality and signal-to-noise criteria. Additionally, we exclusively retain entries that possess Gaia DR3 GSP-Phot temperatures. After this procedure, we have 433,097 entries in the interception of APOGEE and Gaia and 291,065 entries in the interception of GALAH and Gaia. 

\begin{figure}
    \begin{minipage}[ht]{\linewidth}	\center{\includegraphics[width=\textwidth]{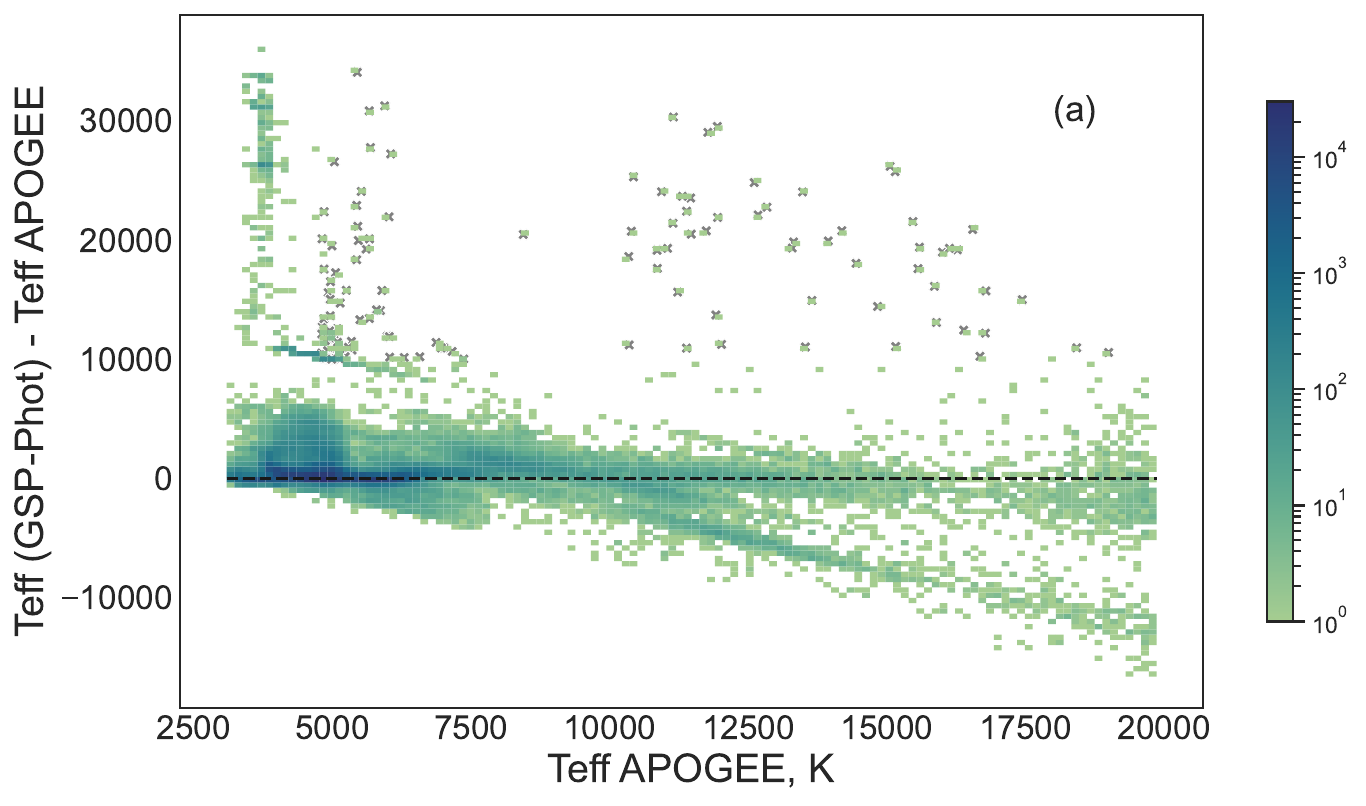}}
	\end{minipage} 
 \\
	\begin{minipage}[ht]{\linewidth}	\center{\includegraphics[width=\textwidth]{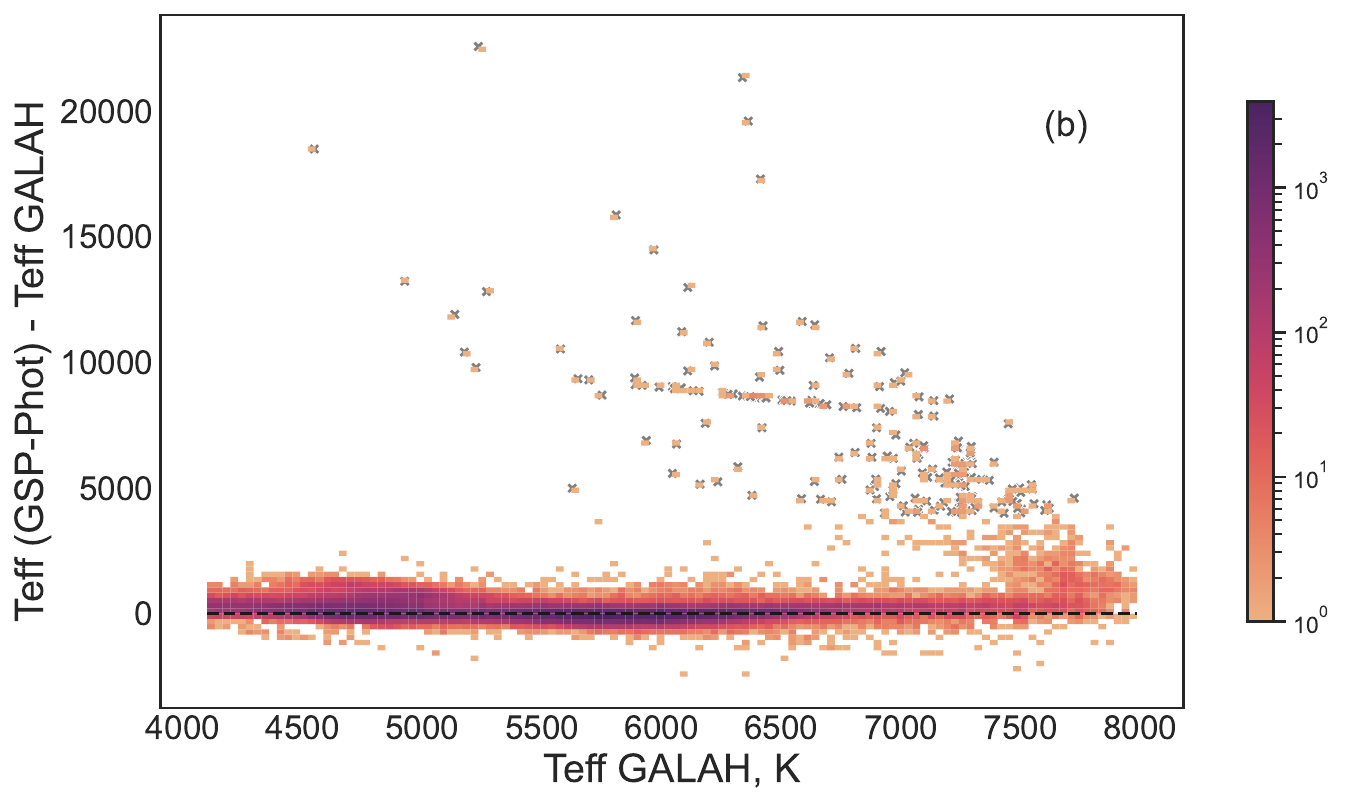}}
	\end{minipage} 
    \caption{Comparison of Gaia DR3 effective temperatures with APOGEE (panel a) and GALAH (panel b) effective temperatures. Colour-code of points is for the stellar density and dashed line is the zero-point.}
    \label{fig:fig1}
\end{figure}

Fig.~\ref{fig:fig1} shows a comparison of Gaia DR3 effective temperatures with the effective temperatures from APOGEE(a)/ GALAH(b) surveys. It represents the discrepancy between the effective temperatures across the effective temperature of a high-resolution survey. A significant difference is observed between the effective temperatures of Gaia and APOGEE. The effective temperatures of Gaia are systematically higher than the effective temperatures of APOGEE. 
For all stars in common after quality cuts between APOGEE and Gaia, a $T_\textrm{eff}$ discrepancy with the median is found.

This difference is found to be non-uniform, displaying a distinct trend where Gaia DR3 tends to overestimate effective temperatures for cooler stars and underestimate them for hotter stars. However, Gaia effective temperatures exhibit remarkable alignment for the majority of GALAH DR3 temperatures, with a minor number of stars showing deviations. This alignment could potentially be attributed to the specific target selection criteria employed by the GALAH survey and by its quality cuts, which will be further discussed later in this section. 

\begin{figure*}
\center{\includegraphics[width=\textwidth]{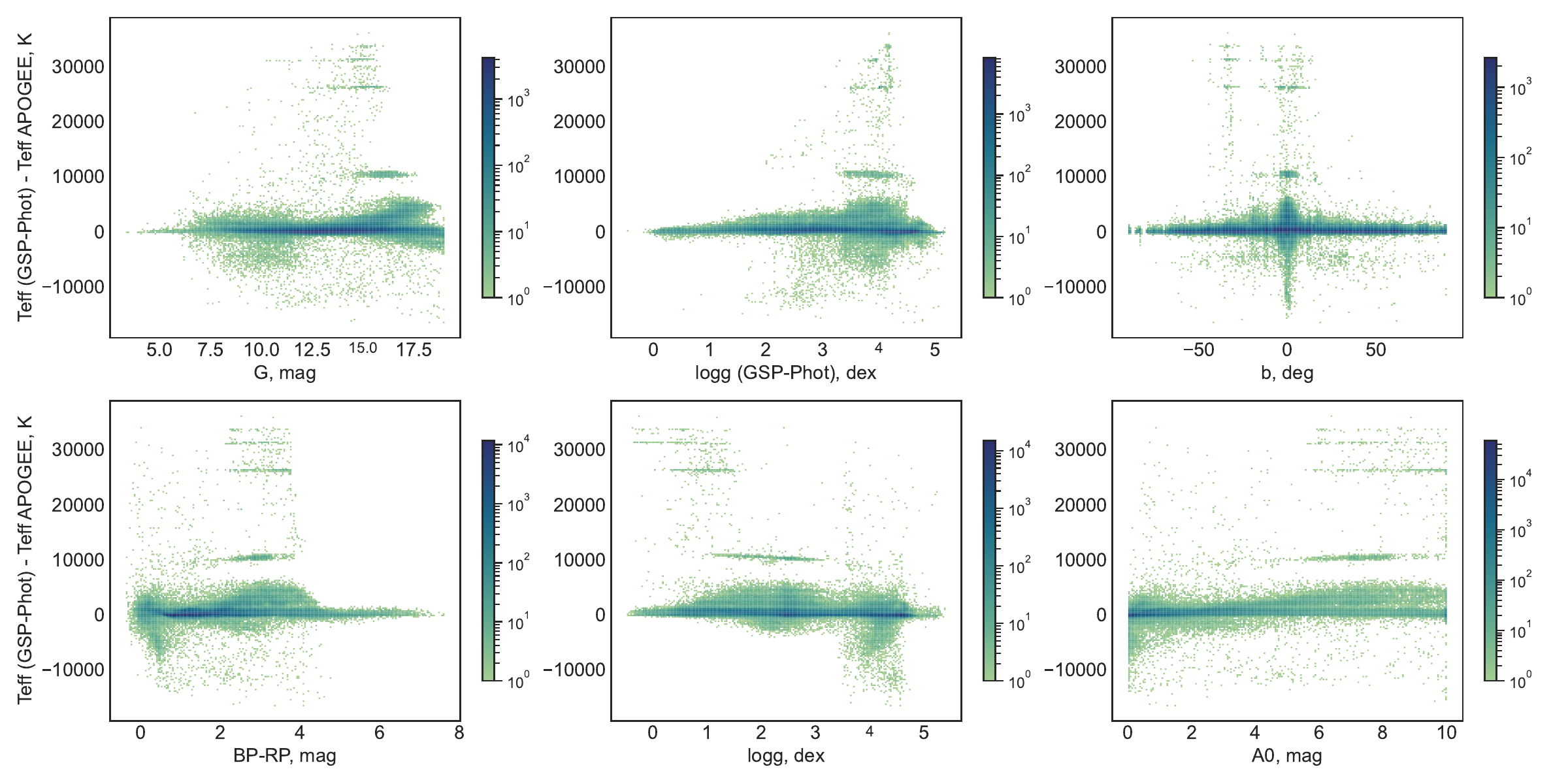}}
    \caption{Difference in effective temperatures between Gaia DR3 and APOGEE as a function of various parameters: $G$ magnitude, $BP-RP$ colour index, $\textrm{log} g$ from Gaia DR3 and APOGEE, galactic latitude $\textrm{b}$, and $\textrm{A}_0$. The data points are colour-coded according to the stellar density.  Details and analysis can be found in the text.}
    \label{fig:fig2}
\end{figure*}

To elucidate the extreme discrepancy observed in effective temperatures between Gaia DR3 and APOGEE, we conduct an in-depth investigation across various parameters. We present the difference as a function of $G$ magnitude, $BP-RP$ colour index, $\textrm{log} g$ provided by Gaia and APOGEE, galactic latitude $\textrm{b}$ and {$\textrm{A}_0$} in  Fig.~\ref{fig:fig2}. 

Some stars within the magnitude range of 12 to 18 mag exhibit a significant discrepancy of up to 35000~K in effective temperatures between Gaia and APOGEE. Possible systematics at the faint end was mentioned in Chapter 11 of Gaia DR3 Documentation \citep{Ulla2022}, stating that discrepancies may occur at $G>16$ mag for $\textrm{log} g$. Conversely, stars with $G<7$ mag demonstrate a notable agreement in effective temperatures across the two surveys. 

Regarding the panels displaying dependability on logarithm of surface gravity ($\textrm{log}g$), it is evident that the large offset in effective temperatures is accompanied by a corresponding offset in $\textrm{log}g$. Notably, stars with the most substantial deviations in $T_\textrm{eff}$ have $\textrm{log}g$ values ranging from 3.3 to 4.4 dex in Gaia DR3, compared to the range of 0-1.5 dex observed for the same stars in APOGEE. 

As it was also noted by Gaia DR3 Documentation \citep{Ulla2022}, in the Galactic plane GSP-Phot module suffers from the temperature-extinction degeneracy, leading to higher differences in astrophysical parameters concerning literature values in this region. Fig.~\ref{fig:fig2} shows that the higher discrepancies are associated with low galactic latitude $\textrm{|b|}<15^\circ$ and high extinction values $\textrm{A}_0>5$~mag. 

The comparison also reveals that the most favourable alignment between Gaia DR3 effective temperatures and APOGEE effective temperatures is evident for stars with a colour index $BP-RP$ greater than 4.5 mag. On the contrary, the poorest alignment is observed for stars with $BP-RP$ values falling within the range of 2-4.5~mag.

The insights gained from the above comparisons provide a basis for understanding why the effective temperatures of Gaia DR3 align better with those from GALAH. The alignment is attributed to the survey's observational strategy, which focuses mainly on nearby stars outside the Galactic plane with a brightness of $G<15$~mag. It naturally eliminates the stars with high extinction, which also exhibit pronounced temperature offsets. This selective filtering results in the removal of a substantial number of poorly calculated effective temperatures in Gaia DR3, contributing to the improved alignment between the two surveys. 

Indeed, while the limits derived from this analysis can aid in refining the astrophysical parameters for the entire Gaia DR3 dataset, it is important to recognize that applying these limits may lead to the elimination of a considerable number of stars with good-quality temperatures, particularly within the Galactic plane. The stringent limits and quality cuts implemented to address issues such as temperature-extinction degeneracy and other systematic errors may unintentionally exclude stars with reliable and accurate temperature estimates.

To overcome this issue of potentially eliminating stars with good-quality temperatures due to stringent cuts, we propose the implementation of a machine learning model trained on the high-resolution spectroscopic surveys discussed in this section. By leveraging the data from these surveys, the machine learning model can learn to identify and account for the systematic errors and uncertainties present in Gaia DR3 temperatures. This approach allows us to avoid the rough use of cuts that might discard valuable data while still achieving accurate and refined temperature estimates.

By using a machine learning model trained on high-quality data, we can improve the reliability and precision of effective temperatures in Gaia DR3, particularly for stars in challenging regions like the Galactic plane. This method offers a more sophisticated and data-driven approach to address the complexities associated with temperature estimation, thereby enabling us to retain a larger subset of stars with good-quality temperatures for comprehensive astronomical analyses.

\section{The data and methods for machine learning approach}

\subsection{The dataset}

The selection of an appropriate training dataset is pivotal for the accurate training of our machine learning model. To ensure the highest level of data quality and robustness, we adopt a training dataset that consists of the intersection of APOGEE and GALAH surveys. The decision to utilize the intersection of these surveys is driven by the desire to leverage the strengths inherent in both datasets. By utilizing their intersection, we aim to enhance the robustness of our training dataset and minimize biases that may be specific to individual surveys. After the quality cuts, mentioned in a previous section, APOGEE DR17 and GALAH DR3 have 17501 stars in common. 

We acknowledge that APOGEE DR17 effective temperatures are calibrated using photometric data, adding an additional layer of accuracy to their temperature estimations. Calibrated effective temperatures are obtained by comparing them with photometric effective temperatures, following the methodology outlined by \citet{2009A&A...497..497G}. In contrast, GALAH DR3, by default, provides temperatures estimated from the best fit of their spectra. This distinction underscores the varied methodologies employed by the surveys, each with its strengths and potential limitations. By incorporating both spectroscopic and photometrically calibrated temperatures within the training dataset, we encompass a broader range of parameter space and ensure a more comprehensive learning experience for our machine-learning model.

Furthermore, to establish a consistent reference for our machine learning model, we calculate a weighted average of the effective temperatures obtained from APOGEE DR17 and GALAH DR3. This reference temperature takes into account the reliability and accuracy of both surveys, allowing for a more balanced representation of the true effective temperatures. This approach also helps mitigate potential biases that could arise from peculiarities within a survey.

However, a constraint of this methodology is the upper limit on the brightness of stars shared between the surveys, which is approximately at a magnitude of $G=15.9$~mag. Additionally, the majority of the stars in the resulting dataset have an effective temperature below 7000~K. As stated in \cite{Joensson2020}, the calibration of the effective temperatures in APOGEE was performed using stars that also primarily fall into that range, so the calibration for the hotter stars might be not well developed. However, we keep all the stars in APOGEE for which the corresponding quality flag is set. All together, these limitations affect the extent to which we can extrapolate the outcomes to encompass stars beyond these defined ranges.

\begin{figure}
\includegraphics[width=0.45\textwidth]{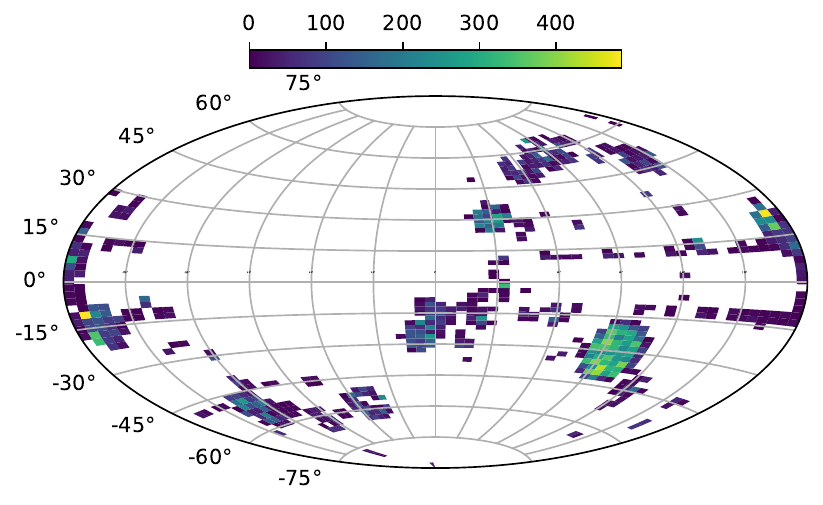}
    \caption{Aitoff projection of the dataset with the Galactic centre at the origin. Points on the map are colour-coded based on stellar density, representing the concentration of stars in different regions. One of the most densely populated regions corresponds to the Large Magellanic Cloud, observable by both GALAH and the southern part of the APOGEE survey, APOGEE-2S.}
    \label{fig:fig3}
\end{figure}

The spatial distribution of objects in the dataset is shown in Fig.~\ref{fig:fig3}. Most points are concentrated within the equatorial plane, with a prominent region aligning with the position of the Large Magellanic Cloud (LMC). Notably, the dataset contains a significant number of points within the Galactic plane. This is particularly significant, as it provides the material for the machine learning model to effectively discern accurate effective temperatures from inaccurate ones in this region. 

\subsubsection{Features}

While Gaia provides an extensive dataset, practical considerations necessitate the selective utilization of its parameters. Therefore, we must carefully choose features that are considered relevant or potentially valuable for evaluating the quality of effective temperatures. We extract information from the following columns of the Gaia DR3 main catalogue and additional astrophysical parameter tables:

\begin{itemize}
    \item \textit{ra} and \textit{dec}: Right ascension and declination coordinates, respectively.
    \item \textit{l} and \textit{b}: Galactic longitude and latitude.
    \item \textit{parallax} and \textit{parallax\_error}: Measure of apparent shift in a star's position due to Earth's motion around the Sun, with the associated uncertainty.
    \item \textit{pm}, \textit{pmra}, and \textit{pmdec}: Proper motion parameters, where \textit{pm} is total proper motion, and \textit{pmra} and \textit{pmdec} are proper motions in right ascension and declination directions, respectively.
    \item \textit{ruwe}: Renormalized Unit Weight Error (RUWE) assessing the reliability of Gaia astrometric solutions.
    \item \textit{ipd\_frac\_multi\_peak} and \textit{ipd\_frac\_odd\_win}: Parameters related to the Integrated Probability Density Function (IPD), describing fractions of windows with multi-peaked and odd-window IPDs. These values contain information about the probability of the source being a double star or being contaminated by another source.
    \item \textit{phot\_g\_mean\_mag}, \textit{bp\_rp}, \textit{bp\_g}, and \textit{g\_rp}: Photometry-related parameters, including mean magnitudes in the G band and colour indices (\textit{bp\_rp}, \textit{bp\_g}, and \textit{g\_rp}) providing information about object colour. We do not use magnitudes in BP and RP bands as they are essentially the linear combination of a magnitude in the G band and a corresponding colour index.    
    \item \textit{teff\_gspphot}, \textit{teff\_gspphot\_lower} and \textit{teff\_gspphot\_upper}: Effective temperature obtained from Gaia's GSP-Phot module with its lower and upper bounds.
    \item \textit{logg\_gspphot}, \textit{logg\_gspphot\_lower} and \textit{logg\_gspphot\_upper}: Surface gravity obtained from Gaia's GSP-Phot module with its lower and upper bounds.
    \item \textit{mh\_gspphot}, \textit{mh\_gspphot\_lower} and \textit{mh\_gspphot\_upper}: Metallicity obtained from Gaia's GSP-Phot module with its lower and upper bounds.
    \item \textit{azero\_gspphot}: Monochromatic extinction ($\textrm{A}_0$) at 541.4 nm, assuming a single-star source, determined by GSP-Phot Aeneas using BP/RP spectra, apparent G magnitude, and parallax.
    \item C*: A modified version of the \textit{bp\_rp\_excess} factor introduced in \cite{Riello2021}. When C* is positive, it signifies that the combined flux from the $BP$ and $RP$ bands exceeds that of the G-band flux, which could imply contamination from nearby sources. Conversely, when C* is negative, it implies the opposite situation, possibly resulting from an over-subtraction of background in either the $BP$ or $RP$ bands
\end{itemize}

The chosen features collectively provide a profile of each star. While our analysis revealed only dependency of temperature differences on galactic latitude (note that it is not entirely symmetrical) we decided to retain longitude as a feature in our model as well. This decision is driven by the well-known large-scale systematics present in Gaia DR3, e.g., see Sec.~3.3 in \citet{2021A&A...649A...5F}. In addition, columns pertaining to contamination and the goodness of the astrometric solution offer insights into the quality of the data. This combined information equips our models with a comprehensive understanding of each object, enabling them to effectively assess temperature quality, identify noteworthy patterns, and account for any data anomalies.

\subsubsection{Classes and imbalance treatment}

We employ a binary classification approach to identify high-quality effective temperatures in Gaia DR3. We use the following selection criterion to partition the data into positive and negative classes:

\begin{equation}
\mid T^{Gaia}_\textrm{eff} - \overline{T}_\textrm{eff}\mid < \delta T^{crit}_\textrm{eff}
\label{eq:eq1}
\end{equation}

\begin{equation}
\overline{T}_\textrm{eff} = \frac{\sum (T_\textrm{eff}^{i}\cdot\frac{1}{\sigma^{2}(T_\textrm{eff}^{i})})}{\sum \frac{1}{\sigma^{2}(T_\textrm{eff}^{i})} }
\label{eq:eq2}
\end{equation}

\begin{figure}
\includegraphics[width=0.45\textwidth]{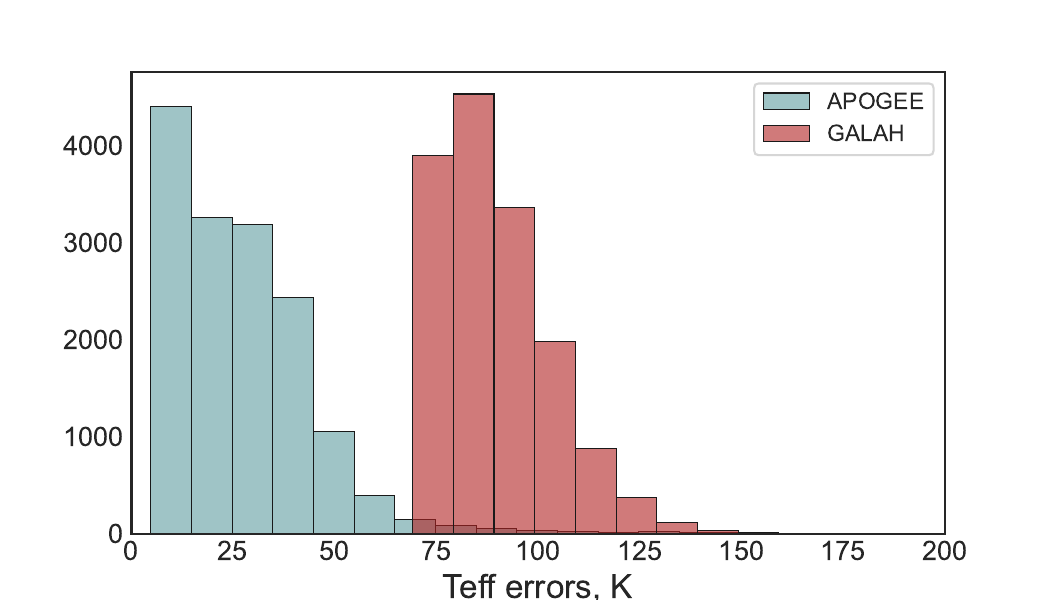}
    \caption{The distribution of the errors provided by surveys, APOGEE and GALAH.}
    \label{fig:fig3.1}
\end{figure}

Here $T^{Gaia}_\textrm{eff}$, $\overline{T}_\textrm{eff}$ are the effective temperature from Gaia DR3 and the weighted average of APOGEE and GALAH surveys, respectively. It should be noted that the errors of the effective temperatures provided in two catalogues are not of the same scale. Fig.~\ref{fig:fig3.1} shows the distribution of the $T_\textrm{eff}$ errors of each survey in their intersection. It can be seen that the errors in APOGEE are significantly lower for all of the objects, which may cause the temperatures preferred by the models to be more inclined towards APOGEE temperatures.

Threshold $\delta T^{crit}_\textrm{eff}$ represents a desired level of accuracy of the result we want to obtain and is flexible. \cite{2023A&A...670A.107H} suggests that the difference between APOGEE and GALAH effective temperatures can be described with a standard deviation of 126.6~K. Confirming that, for effective temperatures below 7000~K, the standard deviation between APOGEE and GALAH effective temperatures is approximately 130~K (likely due to differences in quality cuts). In contrast, the standard deviation for the entire training sample is around 270~K. Although we do not have solid reference temperatures for stars hotter than $\sim 7000~\textrm{K}$, we keep the averaging approach even if the temperatures differ a lot within two surveys.

We use a precision value of 125~K as a desirable threshold and investigate two levels: 125~K and 250~K (hereafter referred to as Threshold-125 and Threshold-250). In each of these cases, the criteria are classified as belonging to the positive class. As pointed out by the reviewer, in this scenario, 125~K is likely the theoretical limit of accuracy for this training dataset.

\begin{table}
    \centering
    \begin{tabular}{c|c|c|c}
\hline
\hline      
         Threshold & $N_{pos}$ & $N_{neg}$ & Percentage of positive\\
         \hline
         125~K & 9497 & 8004 & 54.3\%\\
         250~K & 13707 & 3794 & 78.3\%\\
\hline
\hline         
    \end{tabular}
    \caption{The number of objects of positive (high-quality temperatures) and of negative classes in case of each threshold.}
    \label{tab:datacases}
\end{table}

Tab.~\ref{tab:datacases} shows the number of objects in positive and negative classes in each of the two cases. While the difference between the number of objects in different classes is not substantial in the Threshold-125 case, the dataset is highly imbalanced in the Threshold-250 case. This imbalance primarily stems from the target selection function used by GALAH, as discussed earlier, given that the objects in our dataset also adhere to this selection process.

We use \textsc{smote} \citep{2011arXiv1106.1813C} to oversample minority classes in both cases. It is a resampling technique that generates synthetic samples for the minority class by interpolating between existing samples.  Introducing these synthetic samples aims to create a more balanced class distribution, mitigating the adverse effects of class imbalance on model training.  

The dataset is split into three parts in a stratified fashion: training (60 per cent), validation (20 per cent), and test (20 per cent). We achieve this using the \textsc{train\_test\_split} method from the \textsc{scikit-learn} library. The \textsc{smote} resampling technique is specifically applied to the train portion of the dataset, with the validation and test segments remaining unaltered. Model hyperparameters are chosen based on the validation set using the \textsc{optuna} framework \citep{Akiba2019}. The final evaluation of the model's performance is conducted on the test set.

\subsection{Machine learning models and metrics}

In this study, we investigate three classical machine learning models: XGBoost \citep{Chen2016}, CatBoost \citep{2017arXiv170609516P}, and LightGBM \citep{ke2017lightgbm}. A survey by \citep{2021arXiv211001889B} has highlighted the superior performance of these three boosting algorithms on tabular data.  

Boosting algorithms are a powerful family of machine robust techniques designed to improve the predictive accuracy of models by combining the predictions of multiple weaker models, often referred to as "base learners" or "weak learners". The fundamental idea behind boosting is to sequentially train these base learners, each of whom focuses on the mistakes made by its predecessors. By giving more weight to the examples that are misclassified, boosting algorithms iteratively refine the model's performance, ultimately producing a strong and accurate ensemble model.

We use common classification metrics, namely, precision and recall. Precision quantifies the proportion of relevant instances among those retrieved, while recall measures the effectiveness in capturing relevant instances among all that exist. The definitions are as follows.

\begin{equation*}
    \textsc{Precision} = \frac{\textsc{TP}}{\textsc{TP}+\textsc{FP}} \\
    \textsc{Recall} = \frac{\textsc{TP}}{\textsc{TP}+\textsc{FN}}    
\end{equation*}

True Positives (TP) represent cases correctly identified as positive by the model. True Negatives (TN) are cases correctly identified as negative. False Positives (FP) are  cases wrongly identified as positive when they are negative, and False Negatives (FN) are cases wrongly identified as negative when they are positive. \cite{2019MNRAS.483.5077A} refer to precision and recall metrics as purity and completeness, making them more intuitive. We also calculate the f1 score, a harmonic mean of precision and recall that symmetrically represents both metrics.

It should be noted that due to the target selection strategy employed both by GALAH and APOGEE, the fraction of the positive class in the test part of the dataset could differ significantly from the real-world fraction corresponding to the full Gaia DR3 dataset. This effect is less pronounced in case of Threshold-125, but in case of Threshold-250 the number of positives are much larger than the size of a negative class group. While the recall value should remain unaffected by differences in class imbalance, since it considers only the positive group, precision is more sensitive to such variations. If the fraction of positives is, in fact, larger than in the test sample, then the evaluated precision might be underestimated. Conversely, in case of an underrepresented negative class in the test sample, the precision will be overestimated.

Moreover, when one wants to calculate the conditional probability of the temperature being actually good given that the model has classified it as such, prior probabilities become a pivotal consideration. Consider a scenario in which all temperatures provided by Gaia GSP-Phot are good. In this context, no matter how low precision and recall of the models are, any selection they make would, by default, yield an accurate temperature. Vice versa, when good temperatures are rare, even models with high precision and recall values may yield a low probability that the selected temperature is indeed accurate. This is usually the case when classifying extremely rare classes of objects, such as quasars \citep{2019MNRAS.490.5615B}. We have, at this point, no solid prior probabilities for the Gaia GSP-Phot temperatures. The fractions of good temperatures estimates in Tab.\ref{tab:datacases} are affected by the target selection function of APOGEE and GALAH and may not be fully representative of the broader Gaia GSP-Phot dataset. Nevertheless, these estimates suggest that neither of the two classes, good or bad effective temperatures, is exceedingly rare.

In each model, the hyperparameters play a crucial role in the training process and the complexity of the model. Tab.~\ref{tab:hyp} shows the hyperparameters we have tuned for each model across two threshold cases, denoted as Threshold-125 and Threshold-250. We retained a fixed number of estimators, specifically 500, for both XGBoost and LightGBM models, while other parameters (not mentioned in the table) were employed at their default settings. To do this we have used the \textsc{optuna} framework. For each model in each threshold case, we have created an \textsc{optuna} study aimed at maximizing the f1 score in the validation part of the dataset. 

\begin{table*}
\centering
\caption{Hyperparameters tuned for each model and threshold case. The number of estimators is fixed at 500 for XGBoost and LightGBM models and the number of iterations is fixed at 500 for CatBoost. Other parameters remained at default settings.}
\label{tab:hyperparameters}
\begin{tabular}{|c|c|c|}
\hline
\hline
Model & \multicolumn{2}{c|}{Hyperparameters (Tuned Values)} \\
\cline{2-3}
 & Threshold-125 & Threshold-250 \\
\hline
XGBoost & \begin{tabular}[c]{@{}c@{}}max\_depth (13), learning\_rate (0.014),\\ subsample (0.753),\\ gamma (0.593), reg\_alpha (0.457)\end{tabular} & \begin{tabular}[c]{@{}c@{}}max\_depth (10), learning\_rate (0.061),\\ subsample (0.997),\\ gamma (0.750), reg\_alpha (0.048)\end{tabular} \\
\hline
CatBoost & \begin{tabular}[c]{@{}c@{}}learning\_rate (0.652), depth (5), \\ bootstrap\_type ("Bernoulli"),\\ objective ("Logloss"),  subsample (0.960), \\ boosting\_type("Ordered"), \\  colsample\_bylevel (0.830) \end{tabular} & \begin{tabular}[c]{@{}c@{}}learning\_rate (0.225), depth (11), \\ bootstrap\_type ("Bernoulli"),\\ objective ("CrossEntropy"),  subsample (0.282), \\ boosting\_type("Ordered"), \\  colsample\_bylevel (0.293) \end{tabular}  \\
\hline
LightGBM & \begin{tabular}[c]{@{}c@{}}learning\_rate (0.002), max\_depth (15),\\ num\_leaves (891),  feature\_fraction (0.637) \\ bagging\_fraction (0.808) , bagging\_frec (7) \end{tabular} & \begin{tabular}[c]{@{}c@{}}learning\_rate (0.237), max\_depth (12),\\ num\_leaves (316),  feature\_fraction (0.931) \\ bagging\_fraction (0.959) , bagging\_frec (3) \end{tabular}\\
\hline
\hline
\label{tab:hyp}
\end{tabular}
\end{table*}

\section{Results}

In this section, we present the performance evaluation of our machine learning models on the test subset of the dataset. Additionally, we extend our analysis to assess the models' effectiveness when applied to additional data sources.

\begin{table*}
    \centering
    \caption{Model performance on the test part of the datasets with different thresholds. The best scores are highlighted in bold.}
    \begin{tabular}{lccc|ccc}
        \hline
        \hline
& \multicolumn{3}{c|}{Threshold-125} & \multicolumn{3}{c}{Threshold-250} \\
 & Precision & Recall & F1 Score & Precision & Recall & F1 Score \\
        \hline
        XGBoost & \textbf{0.796} & 0.844 & \textbf{0.819} & \textbf{0.939} & 0.922 & 0.930 \\
        CatBoost & 0.781 & 0.797 & 0.789 & 0.934 & 0.930 & \textbf{0.932} \\
        LightGBM & 0.785 & \textbf{0.845} & 0.814 & 0.926 & \textbf{0.939} & \textbf{0.932}\\
        \hline
        \hline
    \end{tabular}
\label{tab:testscores}    
\end{table*}

Tab.~\ref{tab:testscores} shows the scores achieved on the test subset of the dataset for each model under two cases: Threshold-125 and Threshold-250. It is evident that all the performance metrics, including precision, recall, and f1 score, exhibit significantly higher values in the Threshold-250 scenario. This discrepancy may stem from inherent limitations in the method used to determine effective temperatures in Gaia DR3, even in its optimal conditions. Furthermore, it might be influenced by the initial imbalance present in the Threshold-250 dataset, which persists in the test subset and may contribute to the facilitation of classification. This persistent imbalance in the dataset could potentially result in a more pronounced distinction between positive and negative classes, thereby enhancing the model's performance metrics.

In the Threshold-125 case, precision values are lower than recall values for the same model, indicating that the models have difficulty distinguishing between effective temperatures of different quality, defined by a corresponding 125~K criterion. They tend to classify more cases as positive, including those that are actually negative.  Notably, CatBoost shows a lack of recall in that case, while the difference in precision is smaller compared to XGBoost and LightGBM. 

For the Threshold-250 case, precision and recall are relatively equal, which is desirable in most scenarios. This balance suggests that the model is both accurate in its positive predictions and comprehensive in identifying actual positive cases. Another difference is nearly the same scores for all of the models. This could be an indicator that all models can easily handle the classification with a 250~K threshold, making it more achievable at real-world data.                                                          
 
\subsection{Assessing model performance on additional data sources}

To get a better understanding of how the model will perform outside of the selected training dataset, we evaluate the performance of the model on three additional data sources. These are full APOGEE and GALAH datasets (except for the stars used in training and evaluation of the model), and PASTEL database \citep{2020yCat....102029S}, the bibliographical compilation of stellar atmospheric parameters relying on high-resolution and high signal-to-noise spectroscopy.

From APOGEE and GALAH datasets, we eliminate the objects that have already been used in the training of the models. We apply the same quality cuts as was done in Sec.~\ref{sec:sec2} and exclude stars with missing Gaia DR3 effective temperatures. After that procedure, 451772 objects remain from APOGEE, and 273564 objects remain from GALAH.

\begin{figure}
    \centering
    \includegraphics[width=0.45\textwidth]{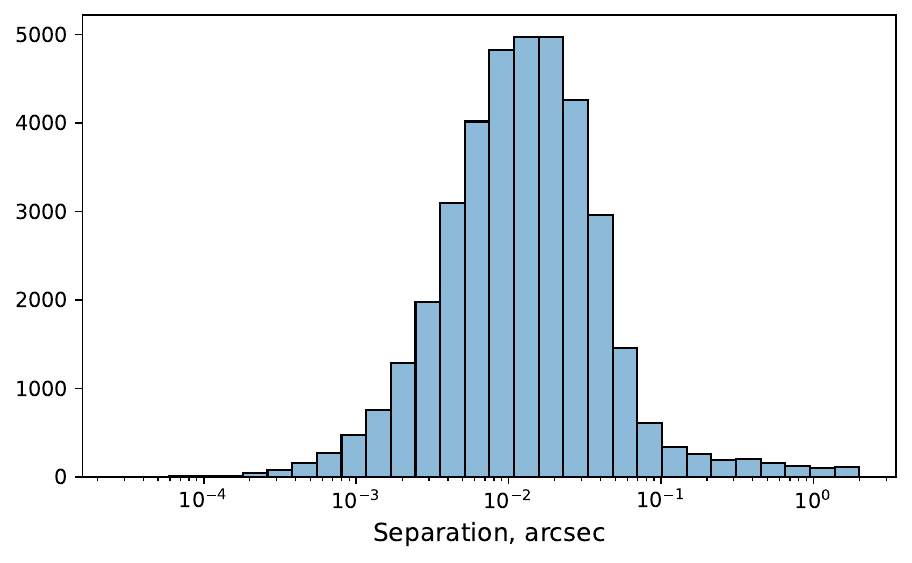}
    \caption{Distribution of angular distance values between the PASTEL objects and best neighbours from Gaia DR3. The majority of the counterparts are found closer than $0.1^{\prime\prime}$.}
    \label{fig:pastel_sep}
\end{figure}

PASTEL database in many cases has several entries for the same object. To avoid ambiguity, we firstly calculate the weighted average of the effective temperatures if there is more than one entry for the object. We then proceed with cross-matching of resulting entries with the Gaia DR3 catalogue. We used the CDS X-Match service with a maximum radius of cross-matching of $2^{\prime\prime}$ to find counterparts for PASTEL entries in Gaia DR3. Fig.~\ref{fig:pastel_sep} shows the angular separation between the PASTEL entries and the counterparts found in Gaia DR3. Best neighbours with valid Gaia DR3 effective temperatures are found within $2^{\prime\prime}$ for 28588 PASTEL objects. 

Objects in these three datasets are labelled in the same fashion as the training dataset: we use two thresholds (125~K and 250~K) to assign objects with either a positive or a negative class. Then we use the corresponding models and extract the objects with high-quality effective temperatures in Gaia DR3 according to each model. We evaluate the results with the same metrics used previously, but we also explore the difference between Gaia DR3 effective temperatures and those from the reference dataset for the objects extracted by the model. It should be noted that while all of these reference datasets are sources of high-quality atmospheric parameters, the values in these datasets can systematically differ from each other as well. 

The results of applying our machine learning models to the datasets are summarized in Tab.~\ref{tab:treshold125} for the Threshold-125 case and in Tab.~\ref{tab:treshold250} for the Threshold-250 case. These tables present an overview of the models' performance across various scenarios, including precision, recall and f1 score metrics. Additionally, we provide models' assessment by calculating the median of the difference and 90th percentile of the absolute difference between Gaia DR3 effective temperatures and the effective temperatures from the reference dataset for the objects extracted by each model. These statistics offer valuable insights into the models' ability to accurately identify high-quality effective temperatures within Gaia DR3. We also add the median and 90th percentile for each full dataset for comparison.

\begin{table*}
\begin{center}
\caption{Performance of machine learning models in the Threshold-125 case. Precision, recall, and f1 score metrics are reported for each model. Additionally, the median of the difference and 90th percentile of the absolute differences between Gaia DR3 effective temperatures and those from the reference dataset are presented for objects classified as high-quality temperatures by each model. Median and 90th percentile values for the entire dataset are included for comparison under the "No model" caption.}
\begin{tabular}{c|ccccc}

\hline
\hline

Model     & Precision & Recall & F1 score  & Median & 90th percentile     \\
\hline
\hline

 \multicolumn{6}{c}{APOGEE}  \\
\hline
No model  & - & - & - & 110.4 & 849.1 \\
XGBoost & 0.778 & 0.768 & 0.773  & -2.3 & 139.8 \\
CatBoost & 0.766 & 0.643 & 0.699  & 1.6 & 151.6 \\
LightGBM & 0.760 & 0.776 & 0.768  & 1.0 & 156.0 \\

\hline
\hline

 \multicolumn{6}{c}{GALAH}  \\
\hline
No model  & - & - & -  & 22.9 & 391.0 \\
XGBoost & 0.710 & 0.674 &  0.692 & -27.0 & 204.3 \\
CatBoost & 0.691 & 0.674 &  0.683 & -16.5 & 212.7 \\
LightGBM & 0.710 & 0.684 & 0.697 & -27.2 & 204.7  \\

\hline
\hline

 \multicolumn{6}{c}{PASTEL}  \\
\hline
No model  & - & - & -  & -64.3 & 524.0 \\
XGBoost & 0.609 & 0.584 & 0.596 & -91.8 &  245.8 \\
CatBoost & 0.600 & 0.497 & 0.544  & -92.8 & 248.9  \\
LightGBM & 0.608 & 0.580 &  0.593 & -89.3 & 242.7  \\

\hline
\hline

\end{tabular}
\label{tab:treshold125}
\end{center}
\end{table*}

\begin{table*}
\begin{center}
\caption{The same table as for Tab.~\ref{tab:treshold125}, but for the Threshold-250 case.}
\begin{tabular}{c|ccccc}

\hline
\hline

Model     & Precision & Recall & F1 score &  Median & 90th percentile      \\
\hline
\hline

 \multicolumn{6}{c}{APOGEE}  \\
\hline
No model  & - & - & -  & 110.4 & 849.1 \\
XGBoost & 0.891 & 0.850  & 0.870 & 15.9 & 260.6\\
CatBoost & 0.872 & 0.887 & 0.879 & 22.4 & 285.2 \\
LightGBM & 0.867 & 0.875 & 0.871  & 22.7 & 294.9 \\

\hline
\hline

 \multicolumn{6}{c}{GALAH}  \\
\hline
No model  & - & - & - &  22.9 & 391.0 \\
XGBoost & 0.907 & 0.821 & 0.862 & -10.5 & 244.4  \\
CatBoost & 0.900 & 0.850 & 0.874 & -6.3 & 250.1 \\
LightGBM & 0.900 & 0.827 & 0.862 & -8.5 & 249.8 \\

\hline
\hline

 \multicolumn{6}{c}{PASTEL}  \\
\hline
No model & - & - & - &  -64.3 & 524.0 \\
XGBoost & 0.875 & 0.865 & 0.870 &  -79.7 & 274.0 \\
CatBoost & 0.872 & 0.872 & 0.872 & -77.6 & 278.2 \\
LightGBM & 0.873 & 0.829 & 0.851 &  -79.8 & 275.9 \\

\hline
\hline

\end{tabular}
\label{tab:treshold250}
\end{center}
\end{table*}

In the Threshold-125 case, our models exhibit relatively low precision, recall, and f1 score. The precision tends to be higher when the models are applied to the APOGEE dataset. The median temperature difference on this data is relatively close to zero and the 90th percentile of the absolute difference is noteworthy, although achieving the desired 125~K threshold was only reached for approximately 77 per cent of the recovered objects. This tendency is most likely due to the disparities in the errors of effective temperatures between GALAH and APOGEE, as illustrated in Fig.~\ref{fig:fig3.1}. Given that APOGEE provided errors are consistently lower, the mean values utilized during training tend to align more closely with the effective temperatures in APOGEE.

When we apply the models to the PASTEL dataset, the median absolute difference shifts towards greater offsets. This could be attributed to the heterogeneous nature of the weighted PASTEL effective temperatures (see discussion of the discrepancies between values from different sources in \cite{2016A&A...591A.118S}). However, it is worth noting that the 90th percentile of the absolute difference in effective temperatures decreases by more than two times compared to the full dataset without selection. This indicates that, while there is a shift towards higher median offsets, the models still contribute to improving the overall quality of the considered temperatures.

As discussed in Sec.~\ref{sec:sec2}, GALAH DR3 target selection promotes better temperature convergence. This can be seen from the parameters of the full dataset. However, even in this scenario, all models contribute to enhancing the statistical quality of Gaia DR3 temperatures. On the other hand, the mediocre recall values across all models suggest that models need to discard many good options in their pursuit of achieving the desired quality. Although in some studies, this value is not decisive, the completeness of extraction of good effective temperatures is an important parameter. 

In the Threshold-250 case, the model scores exhibit a notable improvement across all datasets. The desired 250~K threshold is much more easily achievable in this case, although from 9 to 14 per cent of the stars classified as positive by the models exceed this threshold in terms of the absolute difference between their effective temperatures. The recall score, which reflects completeness, are much better than in Threshold-125 case. However, the models are only able to recover up to 89 per cent of well-estimated stars in the respective datasets. This can be partly attributed to the effective temperatures the models were trained on. As it was already noted the majority of the stars in the training dataset have an effective temperature less than 7000~K. Thus, leading the models to be primarily tailored to capture features within that range. Consequently, they may encounter challenges when dealing with hotter stars.

Notably, all of the models in the case of Threshold-250 are not inclined towards APOGEE effective temperatures as it was in Threshold-125 case. Instead, they appear to favour both GALAH and APOGEE temperatures. This preference contributes to less pronounced offsets in the median values when applied to the PASTEL dataset as well, compared to the Threshold-125 case.
Although the 90th percentile values of the absolute difference in effective temperatures are higher than those in the Threshold-125 case, the models with the 250~K threshold are more versatile and better suited for real-world applications. This suggests that they have broader applicability and can provide more reliable results when dealing with a wider range of stars.

\subsection{Examining the stars selected from APOGEE within a parameter space}

\begin{figure*}
\center{\includegraphics[width=\textwidth]{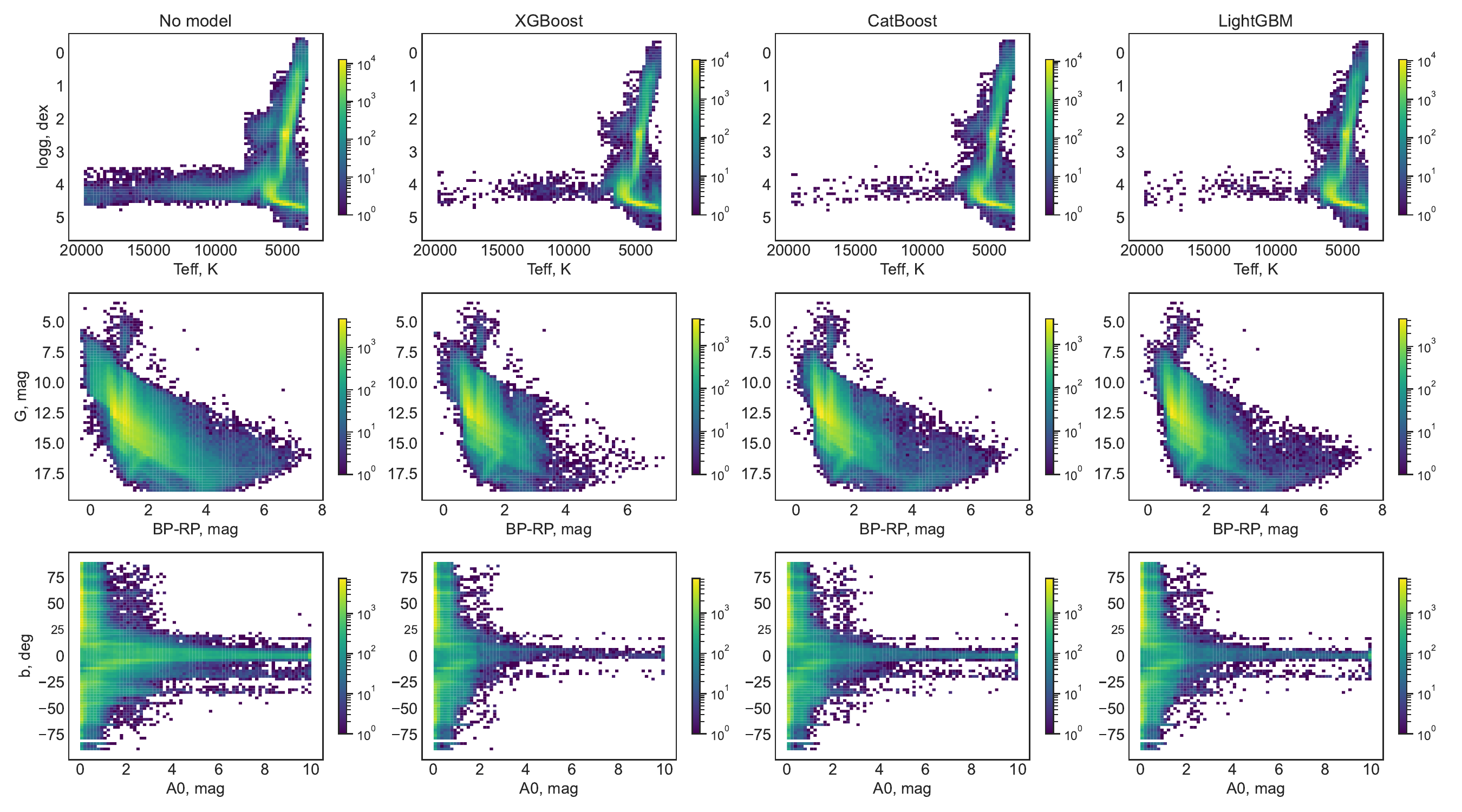}}
    \caption{The distribution of stars from the APOGEE DR17 dataset on three distinct diagrams: $\textrm{log} g$-$T_{\rm eff}$, $G$-$(BP-RP)$, and $\textrm{b}$-$\textrm{A}_0$. The leftmost column shows the complete APOGEE dataset, while the other columns exhibit only the data recognised as good quality by the respective models. The points are coloured according to the stellar density.}
    \label{fig:fig5}
\end{figure*}

One of the motivations for employing machine learning methods was the need to differentiate reliable temperature estimations in demanding regions of the parameter space, including the Galactic plane, areas with high extinction, and regions with faint magnitudes. In this section, we explore the retrieved effective temperatures within this challenging parameter space.

We present (see Fig.~\ref{fig:fig5}) the distribution of APOGEE stars on three diagrams: $\textrm{log} g$-$T_{\rm eff}$, $G$-($BP-RP$), and $\textrm{b}$-$\textrm{A}_0$. We chose to use APOGEE data for this analysis because it covers a broader range in the parameter space and includes a larger number of stars, providing a more comprehensive view of the distribution. The first column shows the entire APOGEE dataset, while the subsequent columns present only the data that the respective model classifies as good quality. All the models applied are in the Threshold-250 version. 

Fig.~\ref{fig:fig5} shows that the applied methods reduce the number of stars but are not selective based on any of the parameters along the axes, with the only exception being hot stars. The models can handle classification even in complex parameter regions such as faint stars and those within the Galactic plane. This suggests the potential for distinguishing reliable effective temperatures in intricate regions of the parameter space.

\subsection{Application to all GSP-Phot effective temperatures}

We produced quality flags for all Gaia DR3 stars with effective temperatures, using XGBoost Threshold-250 model. XGBoost was chosen as the model with minimal among other models 90-th percentile of the absolute difference between reference effective temperatures and the ones selected by the model. The model consider 313 millions of stars of GSP-Phot module to be of good-quality. In this subsection we, at the suggestion of the reviewer, conduct statistical analysis and explore the difference between the effective temperatures with flags 0 (bad temperatures) and 1 (good temperatures) in the parameter space. 

\begin{figure*}
\includegraphics[width=0.7\textwidth]{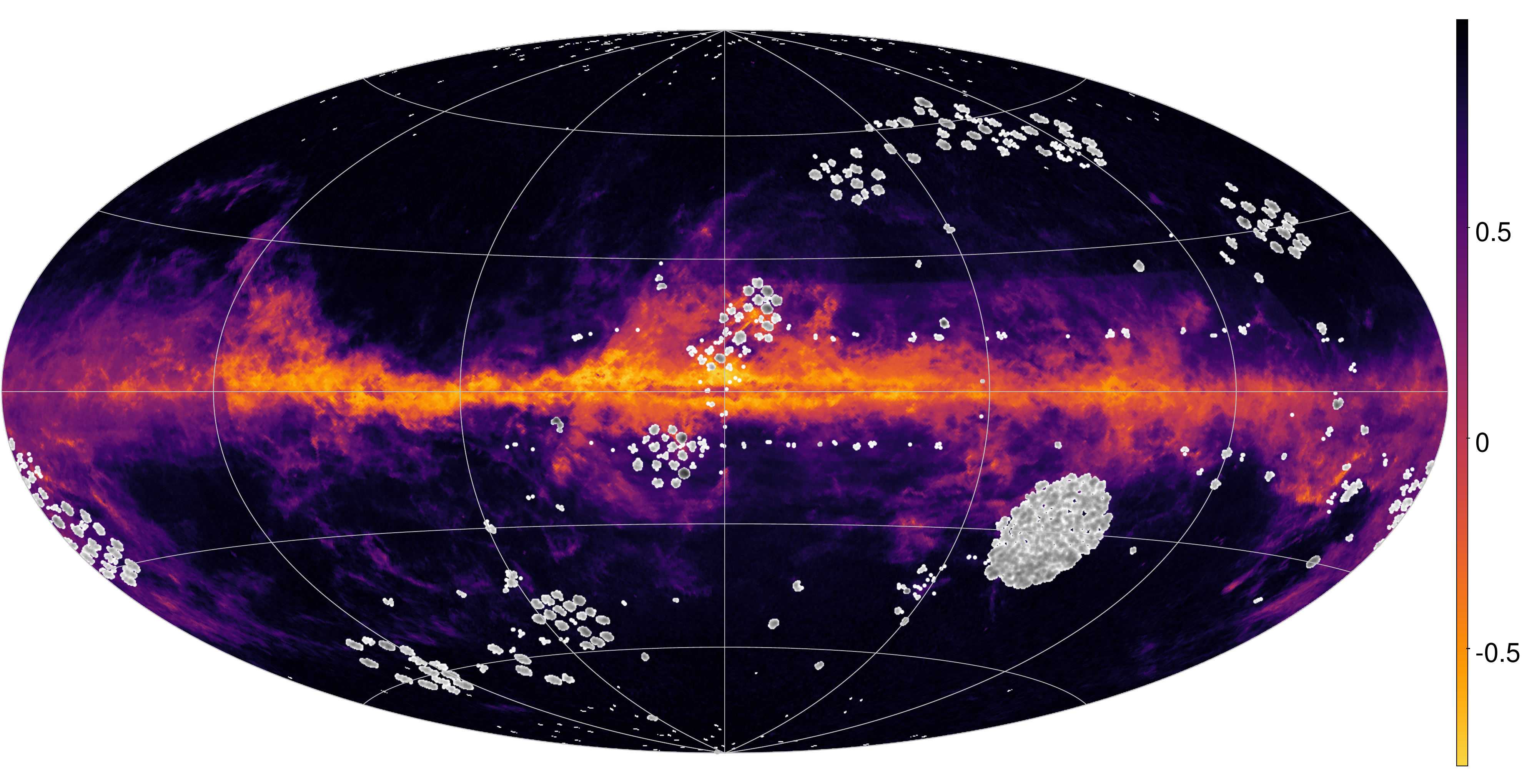}
    \caption{The relative difference between the stellar density of good-quality effective temperatures and bad-quality ones, divided by the stellar density of all objects. In white are footprints of the training dataset. White areas represent the footprints of the training dataset, while bright yellow regions are dominated by bad temperatures, and dark violet regions indicate the presence of good temperatures. The discussion is provided in the text.}
    \label{fig:fig6}
\end{figure*}

Fig.~\ref{fig:fig6} shows the relative difference in stellar density between good-quality and bad-quality effective temperatures predicted by the model. The plot is normalized to the distribution of all Gaia DR3 objects with defined GSP-Phot temperatures.Bright yellow areas indicate regions predominantly characterized by bad effective temperatures. Notably, these regions generally correspond to the distribution of high-extinction areas within the Galaxy. Footprints of the training dataset are superimposed on the plot. Importantly, no discernible correlation is observed between the position of objects in the training dataset and the distribution of good or bad effective temperatures in the results. This observation suggests that the model generalizes well in terms of the spatial distribution across the sky. However, some artefacts are noticeable. Notably, there is an abrupt change in the rate of good temperatures between the Taurus and Cepheus regions on the left side of the plot. A similar border is also evident to the north of the Lupus region, roughly between 240 and 330 degrees of Galactic longitude.   

\begin{figure}
    \begin{minipage}[ht]{\linewidth}	\center{\includegraphics[width=\textwidth]{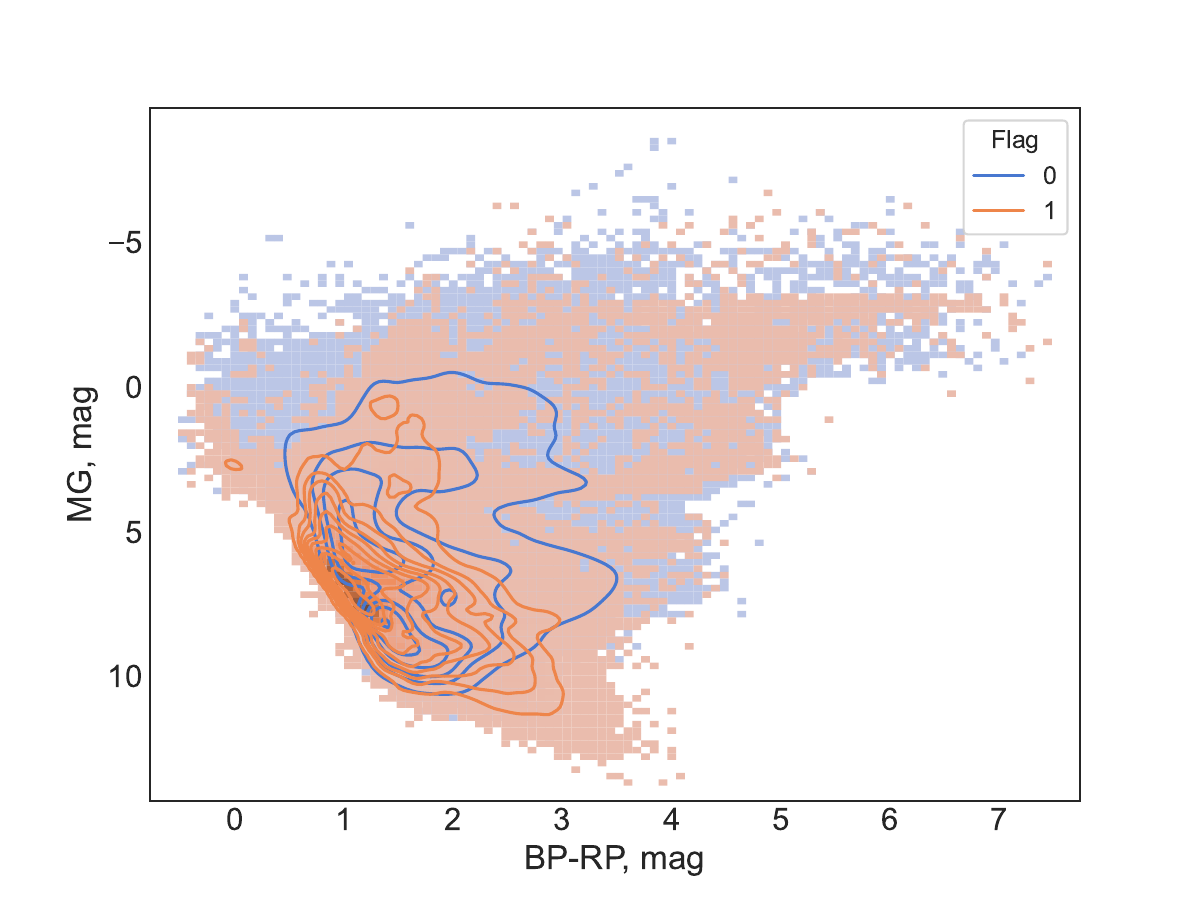}}
	\end{minipage} 
    \caption{The distribution of two distinct subsamples, one with bad effective temperatures (flag 0) and the other with good effective temperatures (flag 1), on the Hertzsprung-Russell diagram.}
    \label{fig:fig7}
\end{figure}

\begin{figure}
    \begin{minipage}[ht]{\linewidth}	\center{\includegraphics[width=\textwidth]{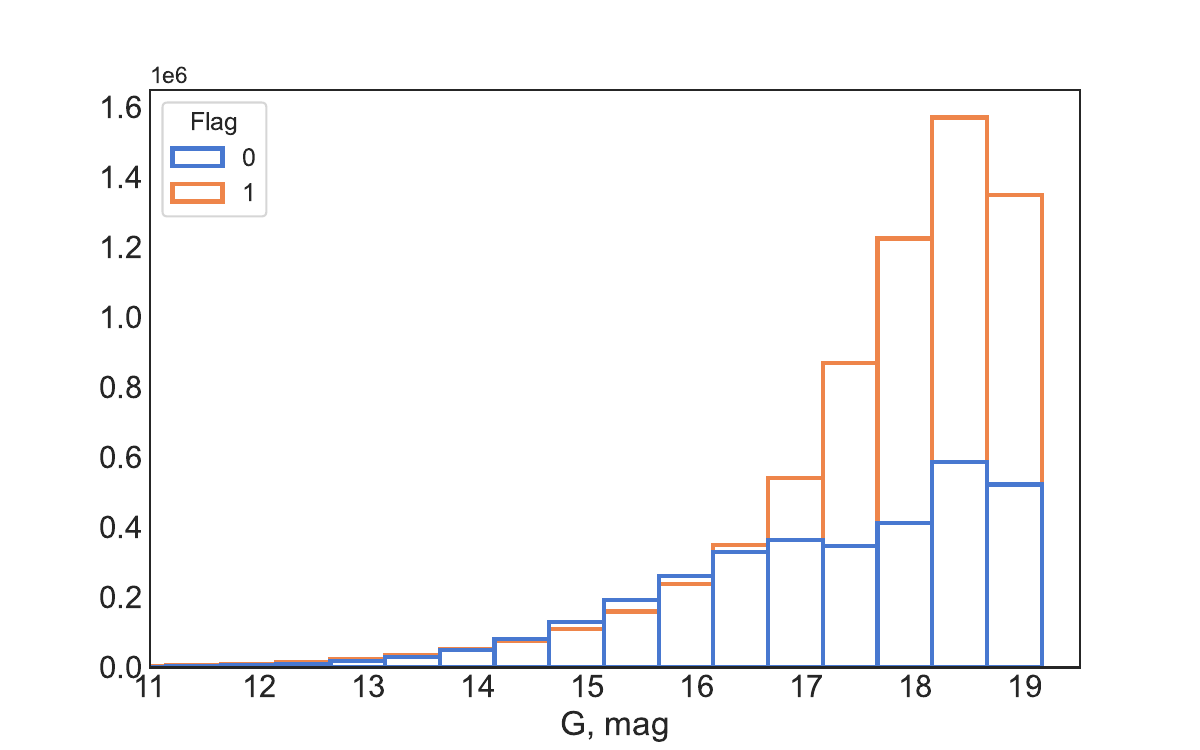}}
	\end{minipage} 
 \\
	\begin{minipage}[ht]{\linewidth}	\center{\includegraphics[width=\textwidth]{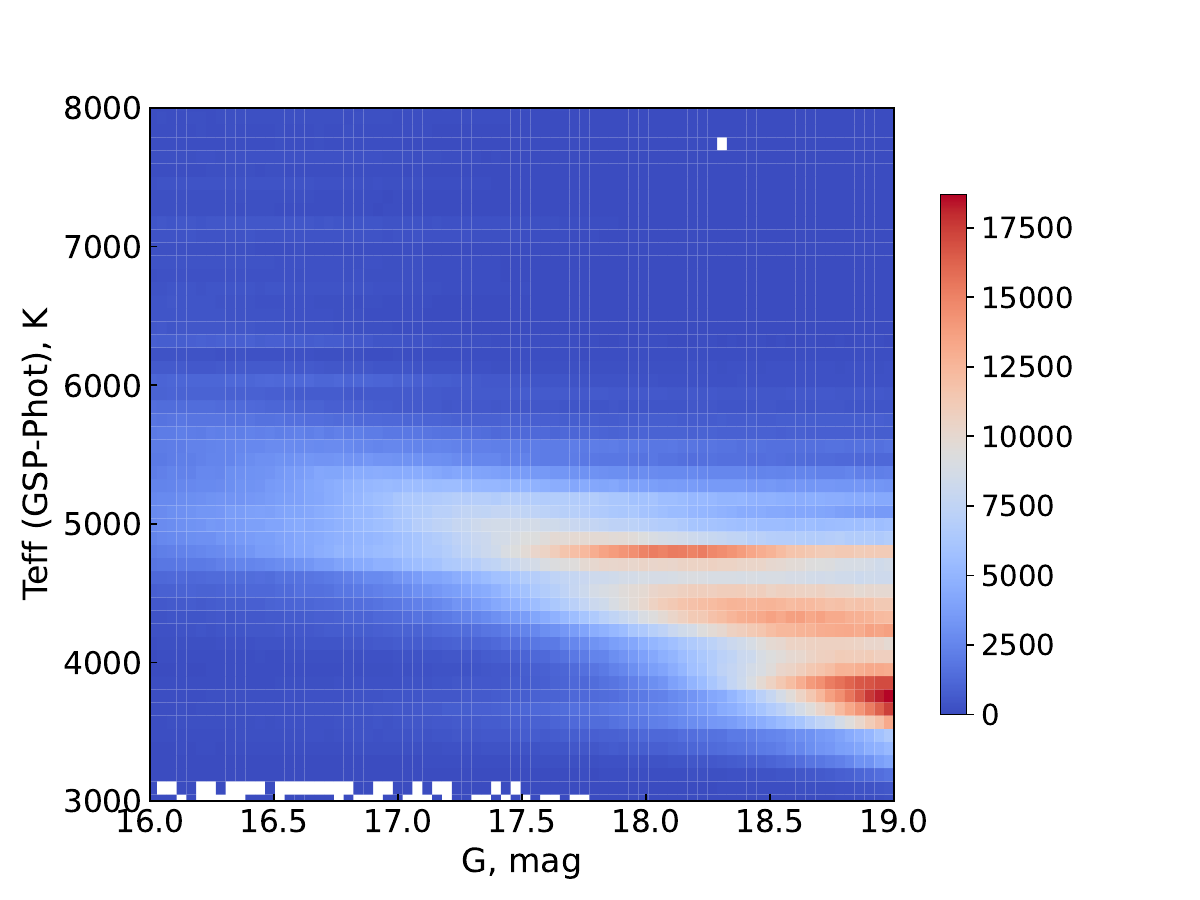}}
	\end{minipage} 
    \caption{The top plot illustrates the distribution of two distinct subsamples, denoted by flag 0 for bad effective temperatures and flag 1 for good effective temperatures, based on apparent magnitude. The bottom plot displays the distribution of Gaia DR3 objects according to both apparent magnitude and effective temperatures. Both plots are generated using a sample of 10 million Gaia objects, and stellar density is represented by color-coding.}
    \label{fig:fig8}
\end{figure}

We compare the distribution of objects with good and bad effective temperatures on the Hertzsprung-Russell diagram in Fig.~\ref{fig:fig7}. This HR diagram was plotted using a sample of one million objects randomly selected from the complete GSP-Phot dataset. Notably, the distribution on the HR diagram exhibits significant disparities. Specifically, objects with good temperatures form a distinct and compact cluster along a well-defined main sequence. In contrast, objects with bad temperatures are more thinly scattered across the plot, implying that the derived absolute magnitudes for this group may be inaccurate. Additionally, it is worth noting that the good-quality sample is devoid of hot stars, and giants are relatively underrepresented, possibly due to the limited number of giants in the training dataset.

Fig.~\ref{fig:fig8} illustrates the G-band apparent magnitude distribution. The top plot categorizes the distribution based on good and bad temperature classifications, while the bottom plot presents the overall distribution by magnitude, disregarding class distinctions. These distributions are derived from a sample of 10 million objects from Gaia DR3. There is a notable rise in the fraction of objects with good temperatures as we observe fainter magnitudes. This elevation in the number of stars, accompanied by a higher proportion of good stars, can be attributed to the emergence of a significant population of stars with GSP-Phot effective temperatures ranging between 3500~K and 5000~K for sources fainter than $G=17.5$ mag, as evident in the bottom plot.

The higher prevalence of good stars among fainter objects is a matter of debate. On one hand, it's crucial to highlight that the training sample employed for the model excludes objects fainter than $G=15.7$ mag. Consequently, the model has limited exposure to stars in this magnitude range. Furthermore, the quality of other parameters, such as astrometric solutions and photometry, tends to decrease with the increasing magnitude. Consequently, this decline in data quality could result in less accurate astrophysical solutions provided by GSP-Phot for fainter stars. Additionally, the temperature-extinction degeneracy might be at play, wherein the extinction for these fainter objects might be underestimated, causing the stars to appear cooler in GSP-Phot estimates. On the flip side, the comparison with APOGEE reveals no specific difference in temperatures for the objects with $G=17.5$ mag and fainter (see Fig.~\ref{fig:fig2}a). For the intersection of APOGEE and Gaia, as described in Sec.~\ref{sec:sec2}, we additionally compared the distribution of the absolute temperature difference for stars brighter than $G=17.5$ mag and for the fainter ones. For each magnitude selection, we compared two GSP-Phot temperature ranges, namely, stars cooler than 5000~K and hotter than that. While the distributions for $G<17.5$ mag are almost equal for cooler and hotter stars, the distributions for $G \geq 17.5$ mag differ significantly between two temperature ranges. We can assume that for stars with $G\geq 17.5$ mag, cooler stars are more prevalent, and they also possess a higher proportion of temperatures that closely align with the APOGEE estimates. Thus, we can neither confirm nor refute the accuracy of the model's results for fainter stars.

\begin{figure}
    \begin{minipage}[ht]{\linewidth}	\center{\includegraphics[width=\textwidth]{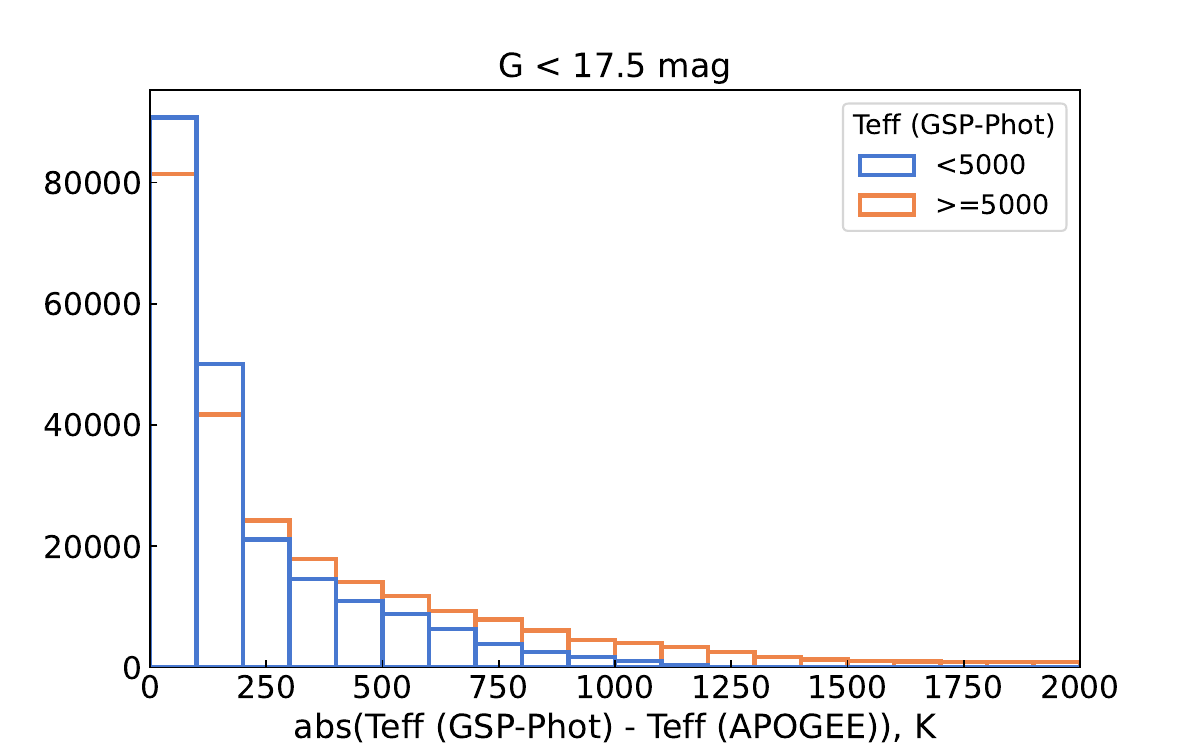}}
	\end{minipage} 
 \\
	\begin{minipage}[ht]{\linewidth}	\center{\includegraphics[width=\textwidth]{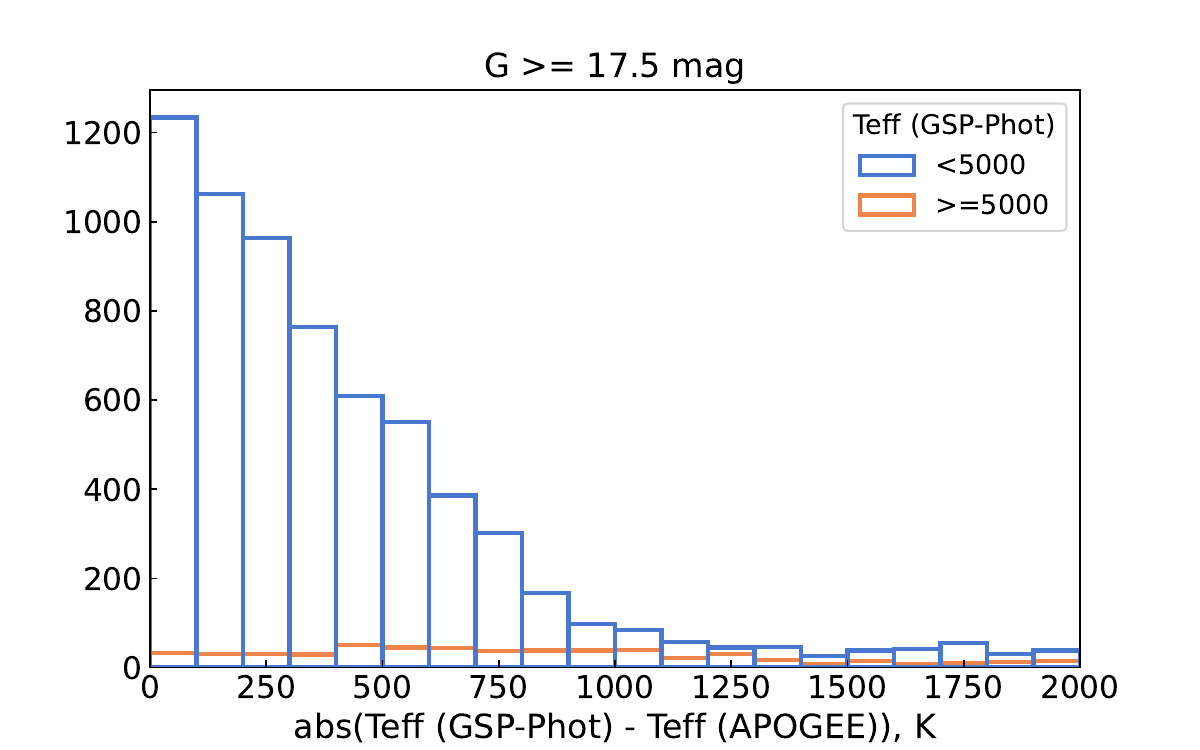}}
	\end{minipage} 
    \caption{The distribution of difference between APOGEE and GSP-Phot effective temperatures for the objects brighter 17 magnitude and fainter than that. Each plot compares the distribution for objects with effective temperatures estimated by GSP-Phot cooler than 5000~K and hotter than that. The x-axis is truncated at 2000~K of temperature discrepancy.}
    \label{fig:fig9}
\end{figure}

\section{Summary and conclusions}

In this study, we compared the effective temperature estimations derived from Gaia Data Release 3 (DR3) with those obtained from high-resolution spectroscopic surveys, specifically APOGEE and GALAH. Furthermore, we explored the application of machine learning techniques to identify good-quality effective temperatures within Gaia DR3 data. One can access all the trained models online for download\footnote{https://github.com/iamaleksandra/Gaia-Teff/}.

The comparison revealed that, while Gaia DR3 provides accurate effective temperature estimations for a considerable number of stars, there is a systematic discrepancy between Gaia DR3's assessments and those from GALAH/APOGEE. In particular, stars close to the Galactic plane with high values of $\textrm{A}_0$ in Gaia DR3 can exhibit offsets in effective temperatures of up to 30000 K. Extremely large offsets are also associated with a particular colour index range, namely, from 2 to 4 mag. Nevertheless, stars with both good-quality and bad-quality temperatures in Gaia DR3 share sometimes the same parameter regions.   

The desire to differentiate between the good-quality and bad-quality temperatures in Gaia, especially in these regions, motivated us to employ machine learning methods. Our objective was to classify the Gaia DR3 temperatures into these two categories, providing a valuable tool for researchers aiming to assess the reliability of effective temperatures for specific objects. 

For establishing a reliable reference for effective temperatures, we used the intersection of APOGEE and GALAH datasets, combining the weighted average of their effective temperatures as a reference value. We examined two definitions of good quality temperature, specifically, $\delta T^{crit}_\textrm{eff}$ equals to 125~K or 250~K as defined in Eq.\ref{eq:eq1}, and applied three boosting algorithms, XGBoost, CatBoost, and LightGBM.  

The results revealed that the Threshold-250 scenario yielded higher performance scores across all models, possibly due to inherent limitations in Gaia DR3's temperature determination. In the Threshold-125 case, precision was lower than recall, suggesting difficulty in distinguishing temperature qualities, while the Threshold-250 case achieved a desirable balance between precision and recall.  

To assess the models' performance on a broader range of data we applied all of the models to three distinct datasets: APOGEE, GALAH and PASTEL. In the case of APOGEE and GALAH, we exclusively considered stars that were not part of the models' training datasets. We found that the precision of temperature estimations in the Threshold-125 case, was relatively low when applying machine learning models to Gaia DR3 data. This indicated challenges in accurately classifying temperatures with the 125~K threshold. The recall score was also low for all of the models in this case, meaning that a lot of good-quality effective temperatures were rejected. In contrast, in the Threshold-250 case, we observed a significant improvement in the model scores across all datasets. The models more effectively achieved the 250~K threshold, making it a more attainable goal. Notably, the models in the Threshold-250 case exhibited a more balanced preference for both GALAH and APOGEE temperature data. 

The distribution of the stars in the parameter space showed that all machine-learning models we use are not constrained by rigid boundaries within the parameter space. Instead, they build a flexible approach, allowing for preserving a significant proportion of objects with good-quality temperatures. This adaptive capacity underscores the potential for using this approach to answer the question if the effective temperature of a particular object is good enough, even within tricky regions.

The limitation of this approach is in the models majorly extract the objects with effective temperatures of less than 7000~K. This is most evidently due to the lack of hotter stars in the training data. Future work is, consequently, in building an approach that allowed for a broader range of temperatures to be extracted with better precision. 

We also produced quality flags for all Gaia DR3 stars with effective temperatures, using XGBoost Threshold-250 model. The dataset with flags is available at \url{https://doi.org/10.5281/zenodo.8325377}. According to the model 313 millions of stars (66 per cent of Gaia DR3 with atmospheric parameters from GSP-Phot module) have good-quality temperatures.

Upon closer examination of the results from the complete GSP-Phot sample, it became evident that the model identified a notably higher number of stars as being of good quality among those with $G \geq 17.5$. This observation raises several potential concerns, such as the model being predominantly trained on a brighter dataset and the existence of the temperature-extinction degeneracy. However, the comparison with APOGEE data reveals that, for fainter stars, temperature convergence is enhanced for cooler stars, which are more abundant at the faint end. This observation lends support to the idea that the model's results for stars with $G \geq 17.5$ may indeed be more accurate. However, it is important to acknowledge the current limitations of this conclusion. At this moment stars with such low magnitude lack reliable temperature references, making it difficult to assess model accuracy for this specific group of stars. Consequently, further research and exploration in this area are warranted.

\section*{Acknowledgements}

The authors express their deep gratitude to the reviewer for their thorough and detailed review,  as well as for providing valuable suggestions that significantly contributed to the improvement of this paper.

Authors thank Elena Pancino for the guidance with Gaia DR3 data and valuable comments on a draft of the paper. Authors also thank Lyudmila Mashonkina and Yury Pakhomov for fruitful discussions. 

OM thanks the CAS President's International Fellowship Initiative (PIFI).

GZ thanks for the support provided by the National Natural Science Foundation of China
under grant nos. 11988101 and 11890694, the National Key R\&D Program of China no. 2019YFA0405500.

This research has made use of NASA's Astrophysics Data System and VizieR catalogue access tool and XMatch cross-matching tool, CDS, Strasbourg, France. This work has made use of data from the European Space Agency (ESA) mission
{\it Gaia} (\url{https://www.cosmos.esa.int/gaia}), processed by the {\it Gaia}
Data Processing and Analysis Consortium (DPAC,
\url{https://www.cosmos.esa.int/web/gaia/dpac/consortium}).

\section*{Data Availability}

The catalogue produced in this work using XGBoost Threshold-250 model
is available for download at \url{https://doi.org/10.5281/zenodo.8325377}. All models considered in this work are available at \url{https://github.com/iamaleksandra/Gaia-Teff}.



\bibliographystyle{mnras}
\bibliography{gaia_teff}   





\bsp	
\label{lastpage}
\end{document}